# Boundary-induced helical bulk acoustic transport in $LiNbO_3$ thin films

Zhe Li[1], Zhen-Hui Qin[1], Shu-Mao Wu[1], Chen-Bei Hao[1], Fan-Yun Pan[1], Hao Yan[1], Yi-Han He[1], Yan-Chen Zhou[1], Xue-Jun Yan[1,2,3], Si-Yuan Yu[1,2,3]*, Cheng He[1,2,3], Ming-Hui Lu[1,2,3], and Yan-Feng Chen[1,2,3]*

[1]National Laboratory of Solid-State Microstructures & Department of Materials Science and Engineering, Nanjing University, Nanjing 210093, China.

[2]Collaborative Innovation Center of Advanced Microstructures, Nanjing University, Nanjing 210093, China

[3]Jiangsu Key Laboratory of Artificial Functional Materials, Nanjing University, Nanjing, 210093, China

*Corresponding author. e-mail: yusiyuan@nju.edu.cn; yfchen@nju.edu.cn

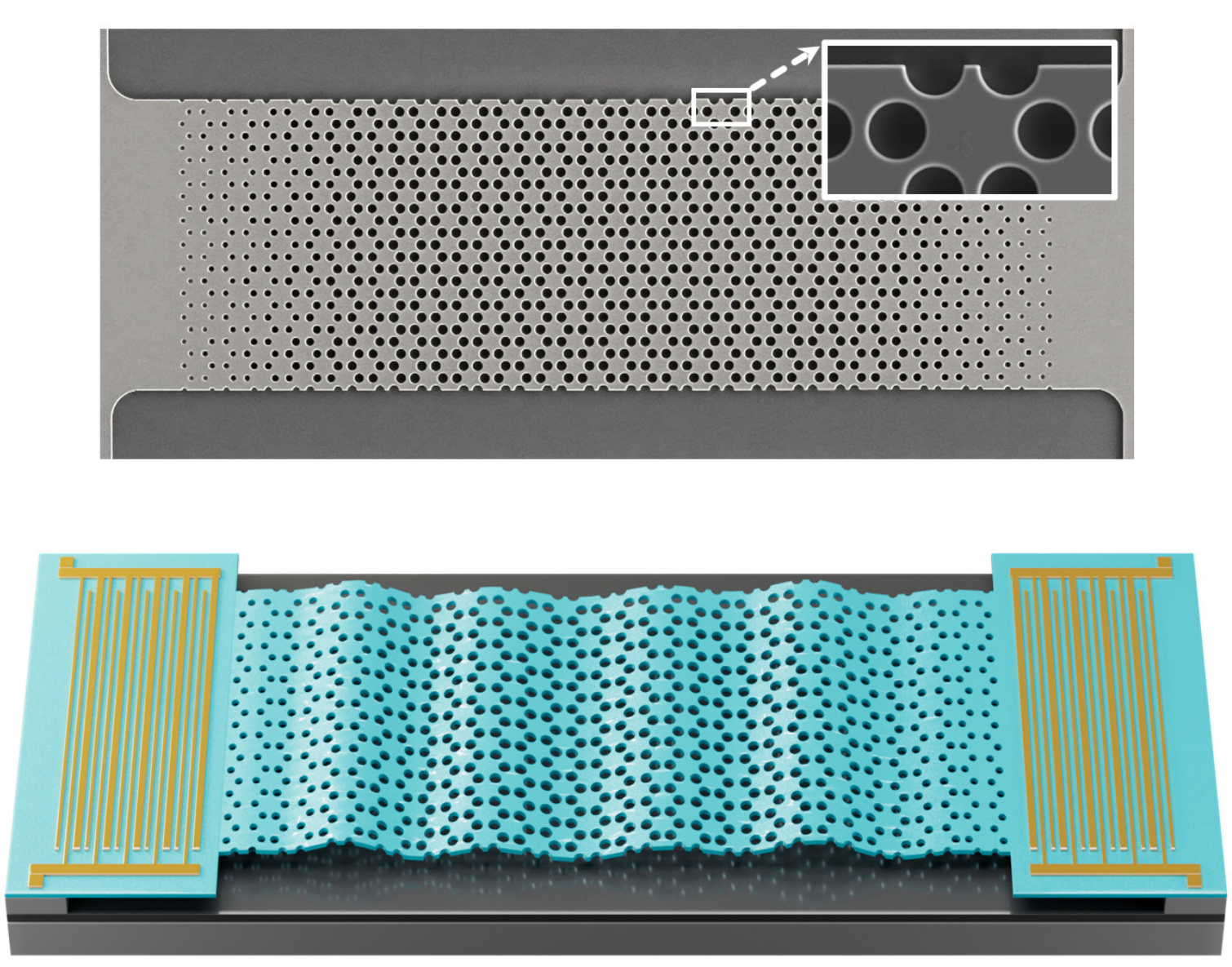

PRL EDITORS' SUGGESTION

Helical, backscattering-resistant acoustic transport in the bulk (not at edges) makes possible the fabrication of efficient, high-throughput on-chip phononic devices.

**We experimentally demonstrate boundary-induced helical bulk states (BI-HBSs) for RF acoustic transport in $LiNbO_3$ thin-film phononic crystals (~175–200 MHz). A boundary-symmetry selection rule at an accidental Γ-point fourfold degeneracy creates interior bulk channels that couple to wide-aperture interdigital transducers without edge-aperture mismatch. Near-field vibrometry and two-port RF S-parameters confirm low-loss propagation with strongly suppressed backscattering through wavelength-scale defects. The helical band also provides slow-wave, low-dispersion delay and phase control on chip.**

## Introduction

On-chip phononics aims to route and process acoustic energy with the same reliability and compactness that photonics brought to light. A persistent obstacle, however, is reconciling backscattering control with efficient, wide-aperture excitation and readout by interdigital transducers (IDTs). Edge-state approaches inspired by topological phases do suppress backscattering [1-11], but they confine transport to narrow interfaces, creating aperture/mode mismatches that limits throughput, degrades signal fidelity, and wastes material utilization in integrated platforms. What is needed is an interior channel that preserves backscattering resistance without giving up full-aperture mode coupling.

Recent macroscopic demonstrations have shown that boundary engineering can select bulk-propagating acoustic channels that retain topology-inspired immunity to disorder [12-14]. Those realizations, largely valley-polarized (quantum valley Hall -like) [2, 4, 15, 16] and implemented at meter-to-table scales with external actuation, confirmed the feasibility of boundary-induced bulk transport yet left two integrated questions open: (i) can such interior channels be realized and quantified on piezoelectric thin films with electrical addressability in the RF regime [17-22], and (ii) can a helical (quantum spin Hall -like) variant—featuring pseudospin–momentum locking under time-reversal symmetry [23-26]—be selected on demand by boundary conditions?

Here we answer both questions by developing a boundary-symmetry design rule tied to a Γ-point symmetry-allowed accidental fourfold (Dirac-like) degeneracy [1, 23-29] in a two-dimensional $LiNbO_3$ phononic crystal (PnC). A minimal tight-binding model links the termination symmetry to the interior channel: choosing traction-free terminations on mirror planes selects helical BI-HBSs that traverse the bulk rather than localizing at edges. We implement this recipe monolithically and electrically excite/measure the channels; near-field laser vibrometry provides spatial field maps while two-port RF characterization yields transmission metrics in a unified, chip-level experiment.

The realized BI-HBSs exhibit three system-level attributes essential for integrated acoustics. First, they maintain strongly suppressed backscattering under wavelength-scale structural defects, demonstrating intrinsic disorder/interface tolerance in an interior geometry. Second, by occupying the bulk aperture they couple ultra-efficiently to wide-aperture broadband IDTs, mitigating the throughput and fidelity bottleneck of edge-confined modes. Third, within the same band they show slow-wave behavior [30-36] with low dispersion [31, 37, 38] enabling precise, compact control of group delay. Together, these results convert boundary-selected interior transport from a macroscopic curiosity into a quantified thin-film piezoelectric integrated platform, providing transferable recipe—Γ-point degeneracy plus boundary-symmetry selection—for robust acoustic transport on chip.

## I. BI-HBSs in suspended $LiNbO_3$ thin films

We engineer a Lamb-wave PnC on a 300-nm-thick Y128°-cut $LiNbO_3$ membrane by etching a honeycomb array of circular holes (lattice constant 7.14 μm, hole diameter 1.97 μm, center offset b = 2.38 μm) using ICP-RIE. The unpatterned film supports the fundamental $A_0$ Lamb branch; the PnC has $C_{6v}$ symmetry and—when restricted to out-of-plane displacement—exhibits a Γ-point symmetry-allowed accidental fourfold (Dirac-like) degeneracy formed by two even $d_{x2-y2}/p_y$-like and two odd $d_{xy}/p_x$-like modes (irreducible representations $E_2 \oplus E_1$; Fig. 1a, b). Here the parity labels denote the sign of the out-of-plane field under reflections about the $x$ and $y$ axes.

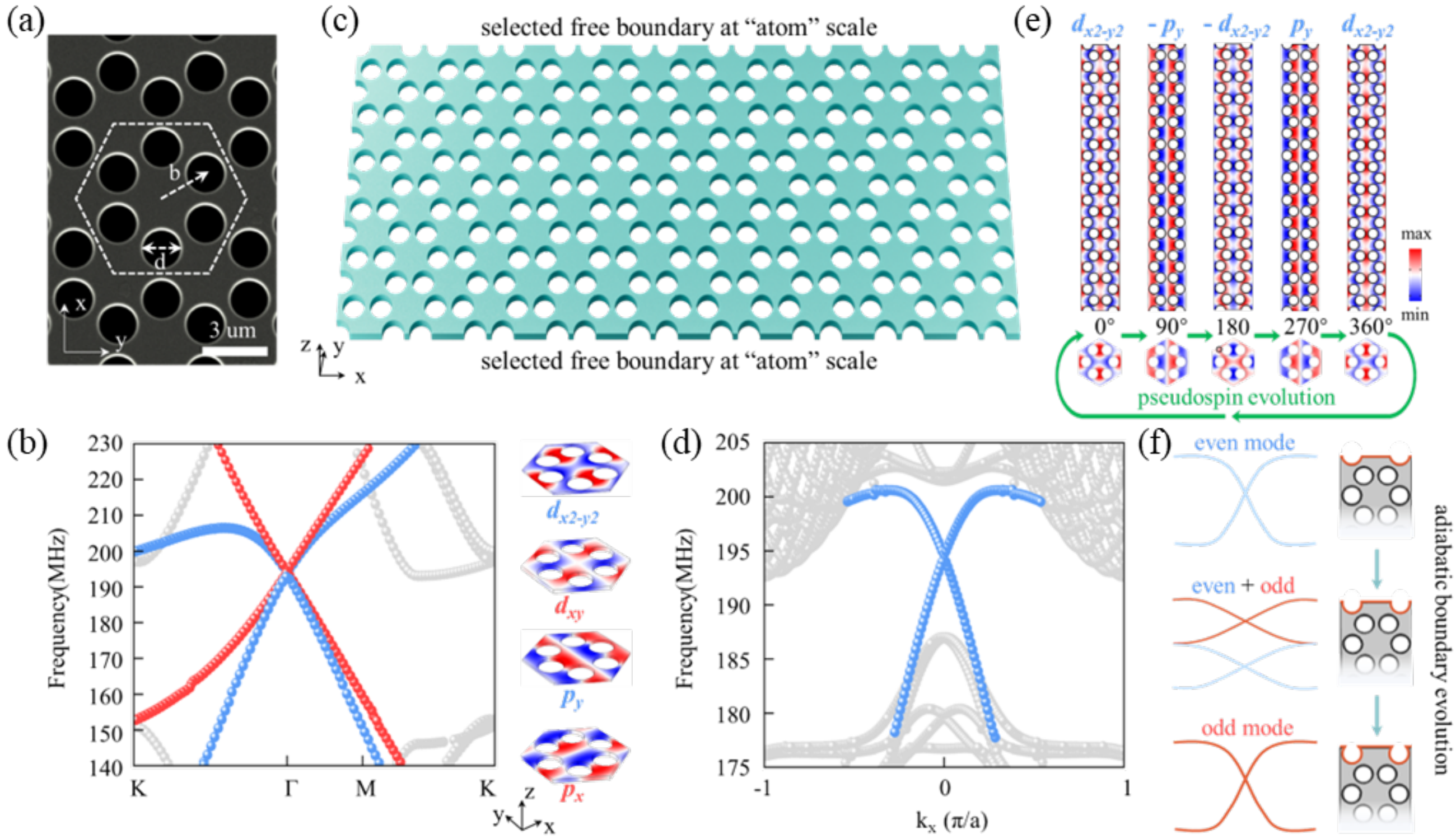

**FIG. 1** Γ-point quadruple degeneracy and boundary-selected BI-HBSs. (**a**) SEM of the honeycomb-lattice $LiNbO_3$ PnC; dashed hexagon marks the unit cell. (**b**) Calculated bands showing a Γ-point symmetry-allowed accidental fourfold (Dirac-like) degeneracy ($E_1 \oplus E_2$); insets: out-of-plane displacement patterns of $d_{x2-y2}$, $d_{xy}$, $p_y$, $p_x$-like Lamb modes. (**c**) PnC ribbon with traction-free terminations along y placed on mirror planes. (**d**) Projected bands for the ribbon, exhibiting interior Dirac-like dispersion dominated by even-parity modes selected by the boundary. (**e**) Out-of-plane fields at different phases, illustrating helical pseudo-spin evolution between $p_y$- and $d_{x2-y2}$- like components. (**f**) Schematic evolution of projected bands as the boundary position is tuned from even- to mixed- to odd-dominant character.

To select interior transport, we terminate the PnC ribbon with traction-free boundaries placed on mirror planes along y (Fig. 1c). Such terminations break the parity balance enforced by the bulk while satisfying zero-traction conditions at the edge; odd-parity combinations cannot simultaneously satisfy mirror-phase continuity and traction-free constraints, leaving even-dominated interior bands across the Brillouin zone (Fig. 1d). The resulting BI-HBSs propagate through the interior rather than along edges;

their pseudo-spin texture manifests as a periodic interconversion between $p_y$- and $d_{x2\text{-}y2}$-like components over the propagation phase (Fig. 1e). A more detailed discussion of the pseudospin is provided in the Supplementary Material (SM [39], Sec. 1). Continuous variation of boundary position tunes the projected spectrum from even-dominated to mixed and eventually odd-dominated character, providing a direct boundary-symmetry knob for selecting interior channels (Fig. 1f, Fig. S3). A compact tight-binding model capturing this boundary-selection rule is provided in the Supplemental Material (SM [39], Sec. 3).

## II. Monolithic BI-HBS transmission lines and electrical metrology

To assess transducer compatibility, we realize "RF–acoustic–RF" transmission lines (TLs) integrating BI-HBS PnC ribbons between single-phase unidirectional transducers (SPUDTs) [52] (Fig. 2a–c). The three-electrode SPUDT (reflector/ground/signal), defined by e-beam evaporation and lift-off, produces unidirectional excitation predominantly into the $A_0$ branch; devices are released by BOE removal of the buried $SiO_2$, yielding suspended membranes. Short impedance-matching sections are placed at the PnC entrances (See SM [39] Sec.6).

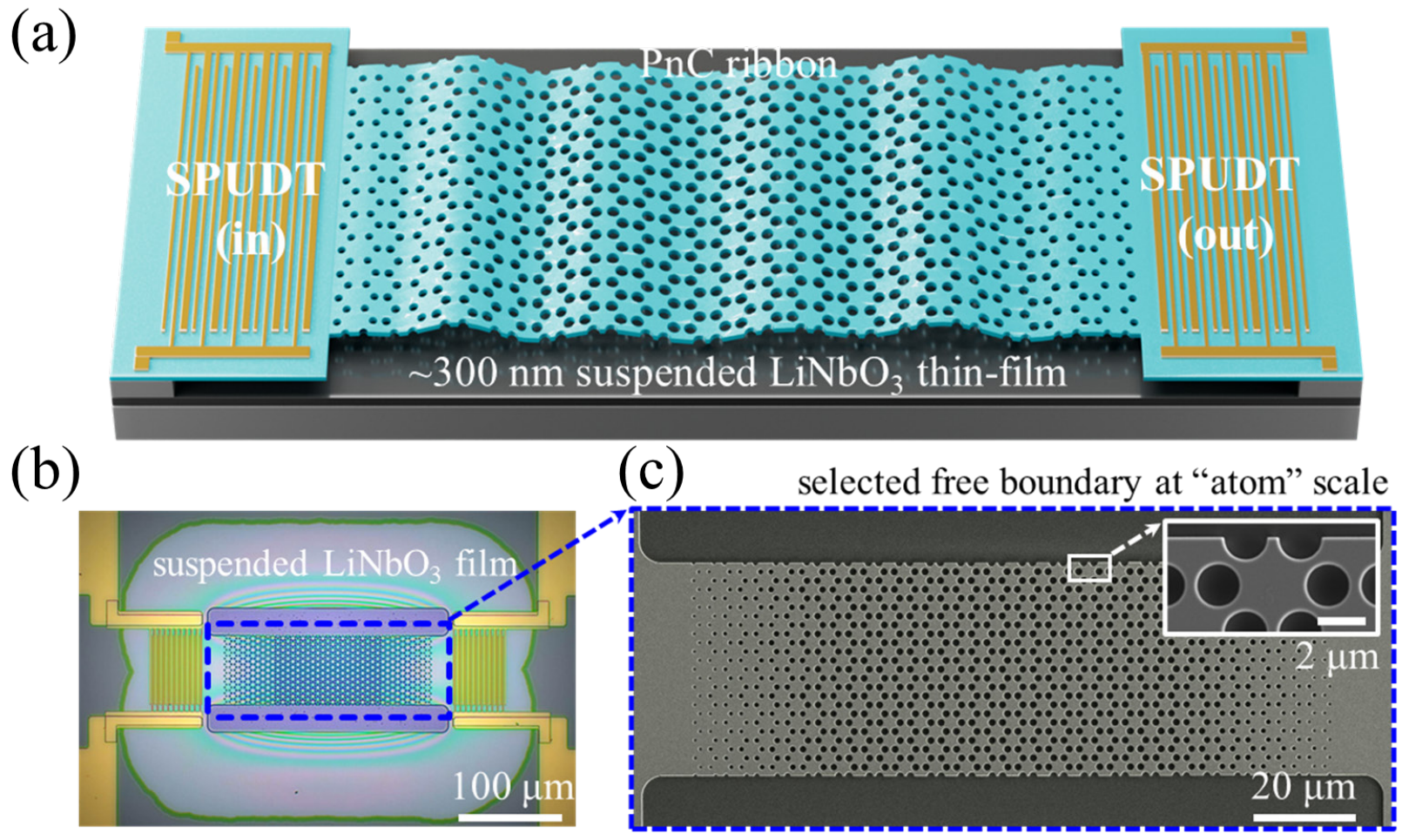

**FIG.2** Monolithic "RF–acoustic–RF" BI-HBS TLs. (**a**) Device schematic: BI-HBS PnC ribbon integrated between SPUDTs on a suspended $LiNbO_3$ membrane. (**b**) Optical micrograph of the released 300-nm-thick membrane. (**c**) SEM of the ribbon highlighting boundary placement and short impedance-matching sections at the entrances/exits (see SM [39] Sec. 6).

We probe the structures with near-field laser vibrometry to map out-of-plane fields and reconstruct projected bands, and with two-port VNA measurements to obtain S-parameters (Fig. 3a). Across ribbons of different apertures (49, 79, 99 μm), the measured field maps show coherent interior modes spanning the full PnC width near ~187 MHz, and the reconstructed dispersions agree with calculations without ad-hoc fitting (Fig. 3b). Electrical transmission $S_{21}$ for BI-HBS TLs exhibits a passband of ≈180–195 MHz (~7%), whereas a uniform-film TL passes a broader 180–210 MHz range (Fig. 3c). Although the shaded region marks the calculated BI-HBS band (~175–200 MHz), the experimentally observed high-throughput window is narrower in practice because of frequency-dependent coupling/excitation efficiency, increased mode mixing near the band edges, and finite-device loss. Reduced $A_0$↔BI-HBS mode conversion toward the high-frequency side further contributes to the sharp roll-off near the upper edge (see SM [39] Sec. 8). Within the BI-HBS band, the insertion loss is comparable to the uniform-film TL, indicating efficient wide-aperture coupling; a sharp roll-off beyond ≈195 MHz (>25 dB) aligns with the calculated band edge. We employ a short impedance matching layer with a gradual hole-diameter taper near the SPUDT–PnC interface to reduce interfacial reflections and enhance coupling. Because the taper is confined to this matching section, it does not appreciably alter the BI-HBS transport band in our operating window, but primarily improves the transmission throughput (see SM [39] Sec. 6). These trends are reproducible across multiple devices and chips measured under identical calibration.

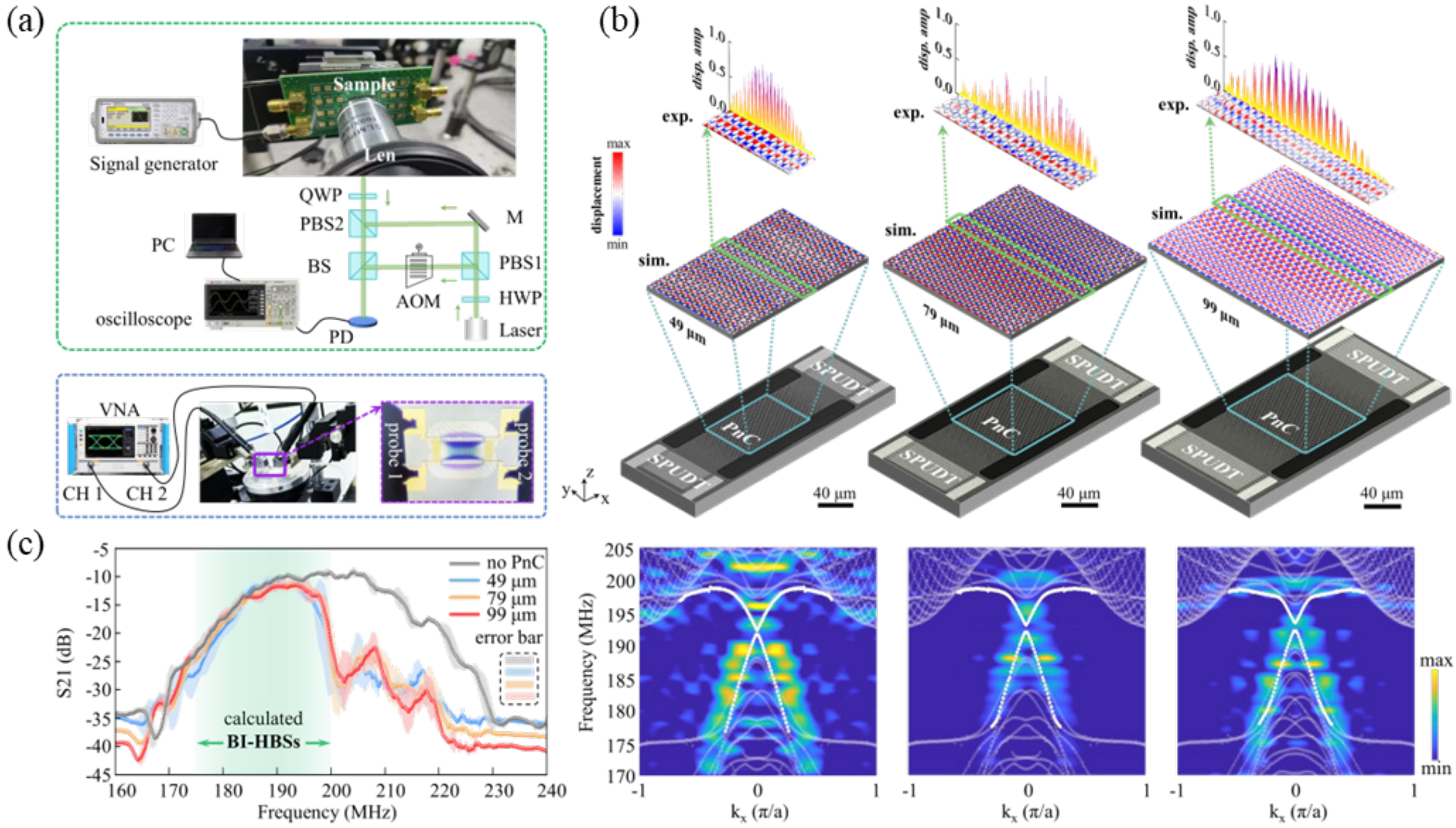

**FIG.3** Experimental verification of BI-HBS transport and transducer compatibility. (**a**) Measurement setup: near-field laser vibrometry and two-port VNA probing with SOLT calibration. (**b**) For three ribbon apertures (49/79/99 μm): measured out-of-plane amplitude maps (top) and displacement fields (second row); simulated fields (third row); SEMs (fourth row); and reconstructed projected bands (bottom) overlaid with calculations (white). Coherent interior modes are observed near 187 MHz with experiment–theory agreement. (**c**) Transmission $S_{21}$ for TLs with and without PnCs: BI-HBS devices show a ≈180–195 MHz (~7%) passband with insertion loss comparable to a uniform-film TL, and a sharp high-frequency roll-off consistent with the BI-HBS band edge.

## III. Backscattering suppression under engineered defects

We next assess backscattering by introducing designed subwavelength defects (holes/slots with controlled sizes and areal densities) into BI-HBS TLs and into reference TLs without PnC (Fig. 4a, b).

Full-field simulations reveal that BI-HBS ribbons maintain uniform energy flow and phase coherence through multiple defects (Fig. 4c, e), whereas the uniform-film TL exhibits pronounced energy depletion and phase disorder (Fig. 4d, f). Experimentally, two-port measurements show that the BI-HBS TL's $S_{21}$ is essentially unchanged across its working band (≈180–195 MHz) after defect insertion, while the uniform-film TL suffers a substantial transmission drop across the transducer band (≈170–230 MHz) (Fig. 4g, h). These results establish backscattering suppression as an intrinsic property of the boundary-selected interior channel. Beyond robustness against defects, the BI-HBS platform suppresses parasitic back-reflections and echo artifacts and enables dispersion engineering together with slow-wave transport, thereby improving phase linearity and group-delay flatness while also supporting, as shown below, enhanced delay per unit length for compact delay/phase-control and echo-sensitive sensing applications [53-55].

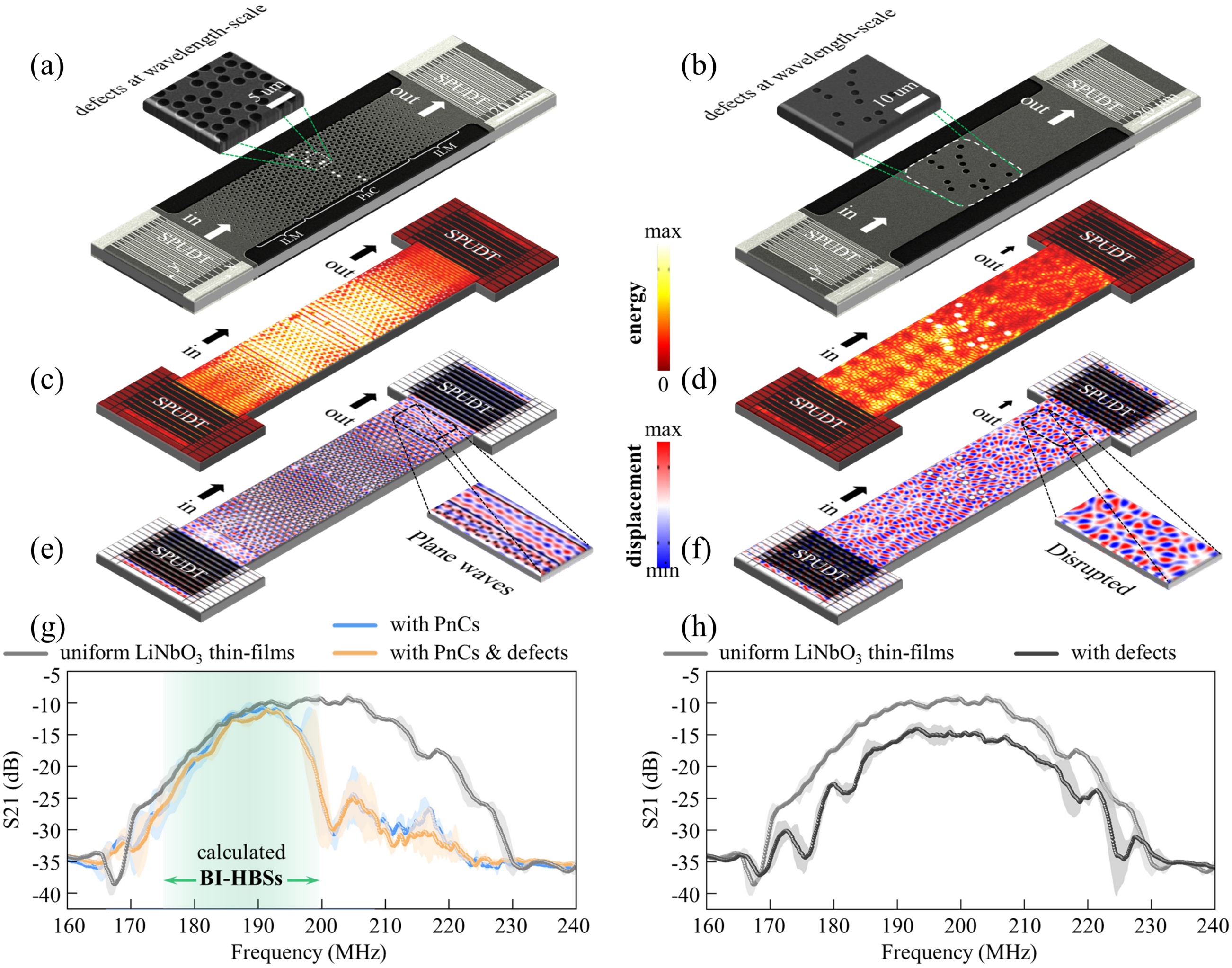

**FIG.4** Backscattering suppression in BI-HBS ribbons under engineered defects. Left: BI-HBS PnC TL; Right: TL without PnC. (**a**,**b**) SEMs showing subwavelength defects (holes/slots with controlled sizes and areal densities). (**c**,**d**) Simulated energy-flux distributions. (**e**,**f**) Simulated out-of-plane displacement fields. BI-HBS ribbons maintain interior transport with phase coherence, while the uniform-film TL exhibits strong disorder-induced disruption. (**g**,**h**) Measured $S_{21}$ before/after defect insertion: the BI-HBS TL remains essentially unchanged within its band, whereas the uniform-film TL degrades across the transducer band.

## IV. Slow-wave BI-HBS delay lines

Finally, we exploit the slow-wave character of BI-HBSs to realize compact delay lines (aperture 49 μm; center-to-center lengths $L$ = 220, 330, 440 μm; Fig. 5a). The length-to-thickness ratio of the PnC region (up to ~1400) highlights the capacity for dense temporal processing. Measured group delay $\tau_g$—extracted from phase dispersion and verified by time-domain gating —increases linearly with $L$ and remains monotonic and continuous within the BI-HBS band, peaking at ~400 ns (Fig. 5b), consistent with weak dispersion. Derived group velocities $v_g = L/\tau_g$ are length-independent and remain in the 600–1100 m s$^{-1}$ range across the band, substantially below the ~1600 m s$^{-1}$ of uniform-film TLs at 187 MHz, thereby yielding an enhanced delay per unit length (Fig. 5c); a brief comparison with state-of-the-art $LiNbO_3$ acoustic delay lines is provided in the Supplemental Material (SM [39], Sec. 9). Importantly, $S_{21}$ spectra of the three BI-HBS delay lines are nearly identical within the working band and comparable to uniform-film TLs (Fig. 5d), demonstrating low-loss slow-wave transport suitable for programmable on-chip delay.

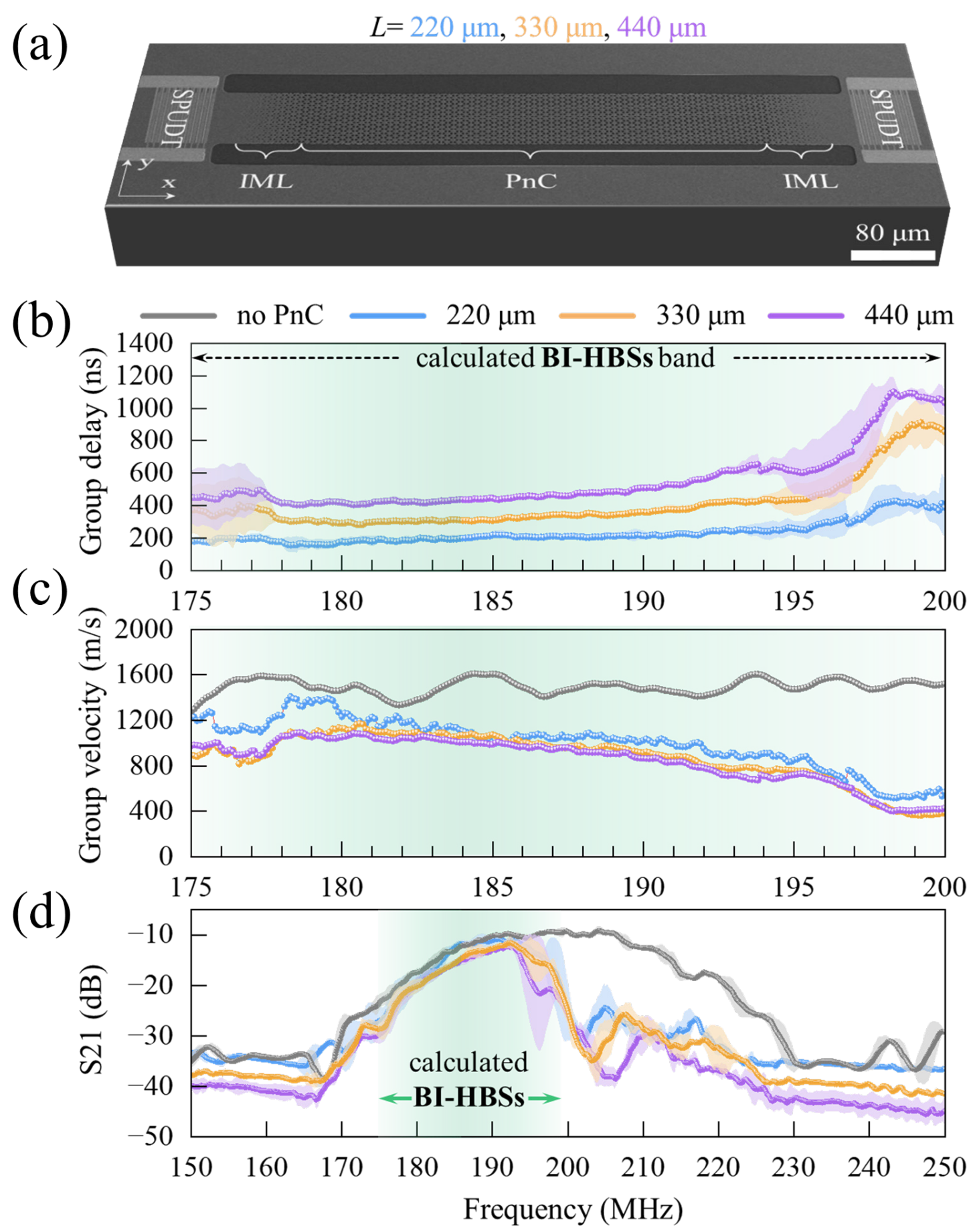

**FIG. 5** Slow-wave BI-HBS delay lines and temporal control. (**a**) Devices with fixed aperture (49 μm) and lengths $L$ = 220 μm, 330 μm, and 440 μm. (**b**) Group delay *v.s.* frequency: delay scales linearly with L and is monotonic within the ≈180–195 MHz BI-HBS band (peak ~400 ns). (**c**) Group velocity extracted as $v_g = L/\tau_g$: 600–1100 m s$^{-1}$ across the band, decreasing with frequency and independent of $L$; the uniform-film TL is faster (~1600 m s$^{-1}$ at 187 MHz). (**d**) $S_{21}$ for the three BI-HBS delay lines: similar passbands and loss, demonstrating low-loss slow-wave transport.

## Conclusion & Discussion

We establish a chip-scale route to BI-HBSs in suspended $LiNbO_3$ thin films, verified by near-field imaging and microwave metrology. A boundary-symmetry selection rule tied to a Γ-point accidental fourfold (Dirac-like) degeneracy selects interior transport without edge confinement, reconciling backscattering suppression with wide-aperture transducer coupling and enabling low-dispersion slow-wave operation. Devices operate at radio frequencies with a ~7% passband and are compatible with standard thin-film processing.

The framework is material-agnostic—set by symmetry and mirror parity rather than substrate specifics—and is readily transferable to $LiTaO_3$, AlN, GaN, and Si/SiN/$SiO_2$ stacks. Immediate directions include bend-tolerance measurements [3, 56], phase-resolved probes of helicity [3, 25], and inverse design to broaden bandwidth and further suppress dispersion [57, 58]. Overall, BI-HBSs furnish a compact, transferable recipe for robust on-chip acoustic transport with wide-aperture electrical access, providing practical building blocks for dense RF routing [20, 59], programmable delay/timing [60, 61], and hybrid photonic–phononic–electronic integration [62, 63].